\documentclass[namedreferences]{solarphysics}
\usepackage[optionalrh]{spr-sola-addons} % For Solar Physics
\usepackage{graphicx}                    % For eps figures, newer & more powerfull
\usepackage{url}                         % For breaking URLs easily trough lines
\newcommand{\apropto}{\;
  \raise0.3ex\hbox{$\propto$\kern-0.75em\raise-1.1ex\hbox{$\sim$
  }}\;\hskip-2pt }

\begin{document}

\begin{article}

\begin{opening}

\title{Occurrence Probability of Large Solar Energetic Particle Events:
 Assessment from Data on Cosmogenic Radionuclides in Lunar Rocks}

%%%%%%%%%%%%%%%%%%%%%%%%%%%%%%%%%%%%%%%%%%%%%%%%%%%
%% Authors Names
%
\author{G.~A.~\surname{Kovaltsov}$^{1}$\sep
    I.~G.~\surname{Usoskin}$^{2,3}$\sep
      }

%%%%%%%%%%%%%%%%%%%%%%%%%%%%%%%%%%%%%%%%%%%%%%%%%%%
%% Runningheads
%
\runningauthor{Kovaltsov \& Usoskin}
\runningtitle{SEP events on long-time scales}

%%%%%%%%%%%%%%%%%%%%%%%%%%%%%%%%%%%%%%%%%%%%%%%%%%%
%% Affilations
%
\institute{$^1$  Ioffe Physical-Technical Institute, 194021 St.Petersburg, Russia\\
            $^{2}$ Sodankyl\"a Geophysical Observatory (Oulu Unit), 90014 University of Oulu, Finland.
                     email: \url{Ilya.Usoskin@oulu.fi}\\
             $^{3}$ Department of Physics, 90014 University of Oulu, Finland.
             }

%%%%%%%%%%%%%%%%%%%%%%%%%%%%%%%%%%%%%%%%%%%%%%%%%%%
%%% Abstract
\begin{abstract}
We revisited assessments of the occurrence probability distribution of large events in solar energetic particles (SEP),
 based on measurements of cosmogenic radionuclides in lunar rocks.
We present a combined cumulative occurrence probability distribution
 of SEP events based on three time scales: directly measured SEP fluences for the last 60 years;
 estimates based on terrestrial cosmogenic radionuclides $^{10}$Be and $^{14}$C for the multi-millennial (Holocene) time scale;
 and cosmogenic radionuclides measured in lunar rocks on the time scale of up to 1 Myr.
All the three time scales yield a consistent distribution.
The data suggest a strong rollover of the occurrence probability so that  SEP events with the fluence
 of protons with energy $>30$ MeV greater than $10^{11}$ (protons cm$^{-2}$ yr$^{-1}$) are not expected at the Myr time scale.

%------------------------------------------------
%--------------Print the Date--------------------
%------------------------------------------------
%setDateLblFont
%30.00 760.00 moveto (Fri Aug 22 19:47:01 2008) show
\end{abstract}

%%%%%%%%%%%%%%%%%%%%%%%%%%%%%%%%%%%%%%%%%%%%%%%%%%%
%% Keywords
%

\end{opening}
%-------------------------------------------------

%%%%%%%%%%%%%%%%%%%%%%%%%%%%%%%%%%%%%%%%%%%%%%%%%%%
%% Sections
%
% \section{}%\label{s:?}

%=====================================
\section{Introduction}

\begin{figure}
\begin{center}
\resizebox{\hsize}{!}{\includegraphics{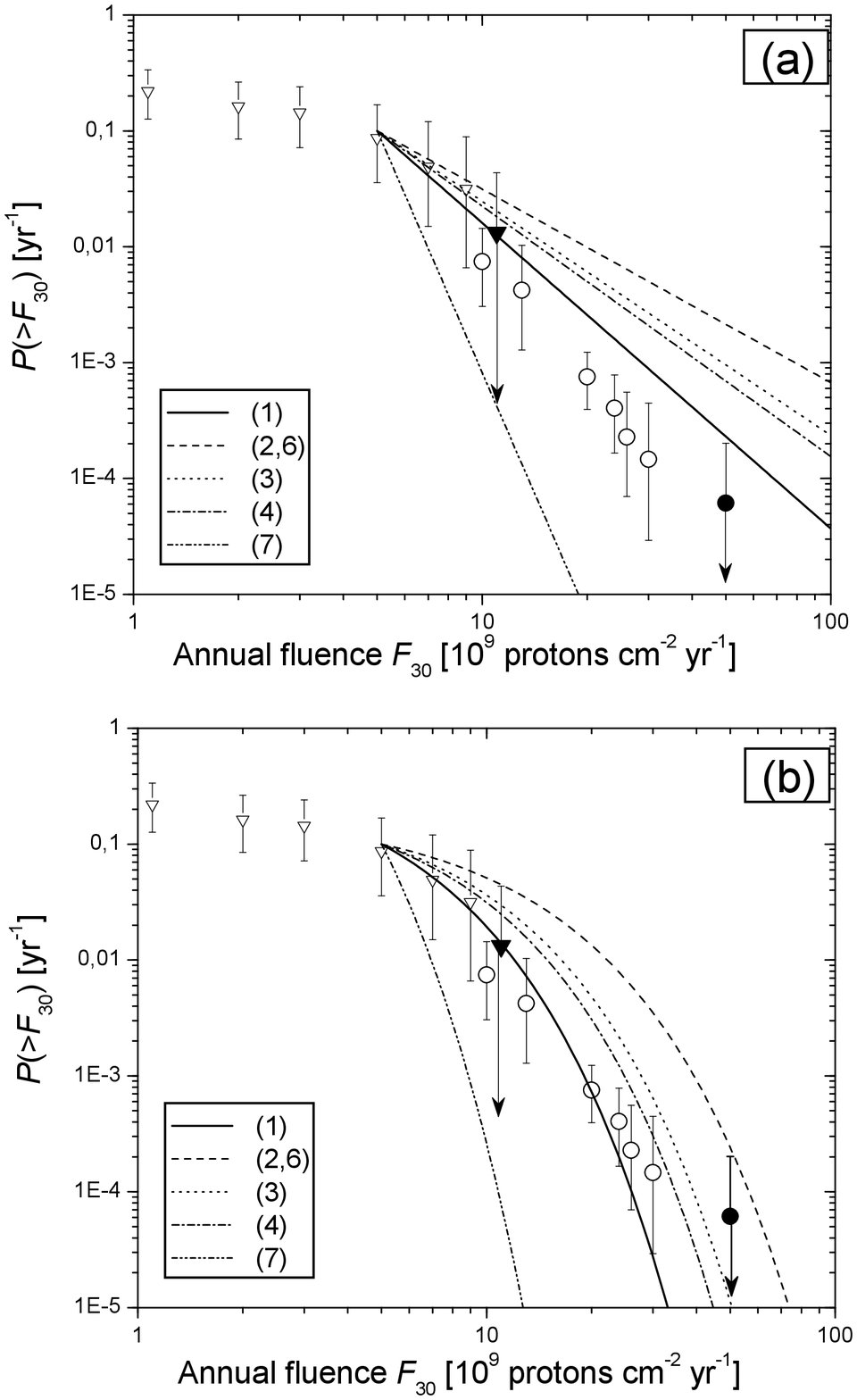}}
\end{center}
\caption{The cumulative OPDF of SEP events (the probability of events with $> 30$ MeV fluence
 greater than the given $F_{30}$ to occur).
 Points with error bars (90\% confidence interval) correspond to the data for the space era since 1955 (triangles) and cosmogenic radionuclides in
 terrestrial archives for the Holocene (circles).
 Open/filled symbols correspond to the measured data and upper estimates, respectively (modified after Usoskin and Kovaltsov, 2012).
 Curves depict best fits of the high-fluence event tail, obtained in this work from lunar data, for two models -- power law [panel (a)]
 and exponential [panel (b)].
 The curves are numbered in the legend, and the numbers correspond to the lines in Table 1.
 All curves converge at the point, corresponding to $P_0=0.1$ and $F_0=5\times 10^9$ protons cm$^{-2}$ yr$^{-1}$.
}
\label{Fig:dist}       % Give a unique label
\end{figure}

Advanced knowledge of the occurrence probability of extreme events related to
 solar energetic particles (SEPs) is very important and acute \cite{hudson10}.
This is important in different aspects: from purely astrophysical questions of the highest possible
 energy released in solar flares \cite{schrijver12} to geo-environment \cite{thomas13}, and even to the
 technological risk assessments \cite{shea12}.
Direct observations of SEPs cover the last six decades with ground-based and space-borne instruments.
The cumulative occurrence probability distribution function (OPDF) for the measured proton ($>30$ MeV)
 annual fluences (\opencite{shea90}; M. Shea, 2012 private communication) is shown in
 Figure~\ref{Fig:dist} as triangles with error bars.
The average annual fluence of SEP ($>30$ MeV) obtained from this data set
 for the period 1955--2007 is $F_{30}=1.1\times 10^{9}$ protons cm$^{-2}$ yr$^{-1}$..
During that period there were four years with $F_{30}$ exceeding $5\times 10^9$ protons cm$^{-2}$ yr$^{-1}$ and no events
 exceeding $10^{10}$ protons cm$^{-2}$ yr$^{-1}$.
The latter makes it possible to obtain an upper limit shown as the filled triangle in Figure~\ref{Fig:dist}.
Most of these strong fluence years were dominated by a single SEP event or a series of consequent events \cite{smart06}.
One can see "steepening" of the OPDF at $F_{30}\approx 5\times 10^9$ protons cm$^{-2}$ yr$^{-1}$, which may indicate that
 stronger events appear more seldom ({\it e.g.}, \opencite{jun07}).
However, the statistics is too low to make any conclusion out of this limited data.
Thus, an extension of the SEP data back in time is needed for a better estimate of the OPDF of strong SEP events.
Such an extension is possible only on a basis of indirect proxies.

One potential proxy was based on nitrate measured in polar ice ({\it e.g.}, \opencite{mccracken01}, \opencite{shea06}), but unfortunately
 it has been shown by \inlinecite{wolff12} that nitrate from Greenland cannot be used as a quantitative proxy for SEP events.
Another potential proxy is related to cosmogenic radionuclides $^{14}$C and $^{10}$Be in terrestrial, independently
 dated archives, where peaks can be associated with strong SEP events \cite{usoskin_GRL_SCR06,usoskin_ApJ_12}.
This method covers the last 10 millennia (the Holocene period) and the corresponding cumulative OPDF is shown
 in Figure~\ref{Fig:dist} as open circles with error bars.
This plot has been updated after Figure~5 of \inlinecite{usoskin_ApJ_12} , by means of combining together
 high- and low-time resolution cosmogenic isotope data and updating the results for the event of 775 AD \cite{usoskin_775_13}.
No events with the annual fluence greater than $5\times 10^{10}$ protons cm$^{-2}$ yr$^{-1}$ have been found, thus
 setting up an upper limit shown as the filled circle in Figure~\ref{Fig:dist}.
However, this method cannot be applied to longer time scales.

An alternative method to evaluate the average flux of SEP on very long time scales is based
 on cosmogenic radionuclides measured in lunar rocks ({\it e.g.}, \opencite{vogt90}).
The method is based on measurements of the depth profile of nuclide's activity in lunar
 rock samples brought to Earth ({\it e.g.}, \opencite{nishiizumi09}).
Standard radionuclides for this method are $^{14}$C (half-life $5.73\times 10^3$ yr), $^{41}$Ca ($1.03\times 10^{5}$ yr),
 $^{81}$Kr ($2.29\times 10^5$ yr), $^{36}$Cl ($3.01\times 10^5$ yr), $^{26}$Al ($7.17\times 10^5$ yr),
 $^{10}$Be ($1.36\times 10^6$ yr), $^{53}$Mn ($3.74\times 10^6$ yr).
However, this method does not have any time resolution, in contrast to the other methods described above, and
 yields only the mean SEP flux integrated over a few life-times of the nuclide.
In particular, it cannot separate SEP events with high fluence from the background of low-fluence events.
Therefore, it is not straightforward to estimate the OPDF for the strong SEP events.
For example, \inlinecite{reedy96} assumed that the entire SEP fluence measured in a lunar rock
 is caused by a single huge SEP event occurred at a half-life of that radionuclide ago.
This is obviously an extreme assumption which gives a very conservative upper limit \cite{reedy96}.
This limit is however not reasonable, since there is always a probability distribution of the events, and a huge event cannot
 appear alone, without a greater number of smaller events occurring.
However, this conservative upper limit has been used quite widely considered as a realistic estimate
 ({\it e.g.}, \opencite{hudson10}, \opencite{schrijver12}).

Here we revise the assessment method for the occurrence probability of SEP events, based on
 cosmogenic radionuclides measured in lunar rocks, and give a more realistic estimate of the OPDF for strong SEP events,
 assuming a rational model for the distribution of the event strengths.

% ======

\section{Modelling}
\label{S:mod}

Let us define the probability of a SEP event with the annual $F_{30}$ fluence exceeding $F$ to occur, as
 $P(F)$.
The mean SEP fluence over a long time period is defined as
\begin{equation}
\langle F\rangle = \int_0^{F_0}{F\cdot p(F)\cdot dF} + \int_{F_0}^\infty{F\cdot p(F)\cdot dF} = \langle F_1\rangle + \langle F_2\rangle,
\label{Eq:Fmean}
\end{equation}
where $p(F)\equiv -dP(F)/dF$ is the differential frequency function for a SEP event with the fluence being exactly $F$.
Here we split the mean fluence into two parts:
 $\langle F_1\rangle$ is the mean fluence defined by low-fluence $(F<F_0)$ but more frequent events, while
  $\langle F_2\rangle$ is due to strong $(F\geq F_0)$ but rarer events.
As a separation we select the annual fluence $F_0=5\times 10^9$ protons cm$^{-2}$ yr$^{-1}$.
From recent instrumental observations continuously conducted since the 1950s
 we estimate that the total $>30$ MeV fluence is $\langle F\rangle =1.1\times 10^9$ protons cm$^{-2}$ yr$^{-1}$, and
 $\langle F_1\rangle=5.2\times 10^8$ protons cm$^{-2}$ yr$^{-1}$, {\it viz.} about half of the total fluence.
The corresponding occurrence probability is $P(F_0)=P_0=0.1$ yr$^{-1}$ (see Figure~\ref{Fig:dist}).

Statistics of the high-fluence events is assessed here using the cosmogenic radionuclide data, measured in lunar samples on the very long time scale.
From such nuclides with different life-times, ranging from millennia to millions of years,
 the mean annual fluence of SEP, $F^*$, was determined based on measurements of their activity in the lunar rocks (see Table~\ref{Table:1} and references therein).
Here we try to estimate, based on these data, what could be the OPDF for rare high-fluence SEP events.
This was done in the following way.
First, the shape of the OPDF tail was {\it a priori} prescribed.
Here we assume two models: power-law and exponentail tails.

%\subsection{Power law probability distribution}
%\label{S:PL}
We first assume that the OPDF has a {\it power-law} shape in the range of high fluences,
  with the upper end being fixed at $P_0$ and $F_0$.
\begin{equation}
P(F)=P_0\left({F\over F_0}\right)^{-\alpha}.
\label{Eq:PL}
\end{equation}
Then $\langle F_2 \rangle$ is directly related to the spectral index $\alpha$ by
\begin{equation}
\langle F_2 \rangle = {\alpha\over \alpha-1}P_0\cdot F_0.
\label{Eq:F_pl}
\end{equation}
%

%\subsection{Exponential probability distribution}
%\label{S:exp}
Next we assume an {\it exponential} OPDF tail for high-fluence events:
\begin{equation}
P(F)=P_0\cdot\exp{(\beta(F_0-F))}.
\label{Eq:EXP}
\end{equation}
Then the $\langle F_2 \rangle$ is directly related to the exponent $\beta$ by
\begin{equation}
\langle F_2 \rangle = P_0\cdot (F_0+{1\over\beta}).
\label{Eq:F_exp}
\end{equation}
%

%\subsection{Monte-Carlo simulations}

For a given value of $\langle F_2\rangle$ one can define, from Equation~\ref{Eq:F_pl} or \ref{Eq:F_exp},
 parameters $\alpha$ or $\beta$, respectively.
However, the uncertainties of thus defined spectral index cannot be straightforwardly calculated,
 and we perform a Monte-Carlo test for each nuclide, characterized by its life (e-folding) time $\tau$.
We made $N$ realizations of the time series, where the occurrence of high-fluence ($F>F_0$) events was
 simulated, at each time $t$ using a random number generator.
First, a random number $R(t)$, corresponding to the year $t$, is picked from the uniform distribution between 0 and 1.
This random number is then converted into the fluence value $F(t)>F_0$ as follows, for the power-law OPDF:
\begin{equation}
R(t)=\int_0^{F(t)}{p\cdot dF}= \int_0^{F_0}{p\cdot dF} + \int_{F_0}^{F(t)}{p\cdot dF}=1-P_0\, F_0\, F(t)^{-\alpha}.
\label{Eq:R}
\end{equation}
Thus, if $R\geq (1-P_0)$
\begin{equation}
F(t)=F_0\, \left({P_0\over 1-R(t)}\right)^{1/\alpha}.
\label{Eq:F_PL}
\end{equation}
For the exponential OPDF, one can similarly obtain
\begin{equation}
F(t)=F_0 - {1\over\beta}\,\ln{\left({1-R(t)\over P_0}\right)}.
\label{Eq:F_exp2}
\end{equation}

\begin{figure}
\begin{center}
\resizebox{\hsize}{!}{\includegraphics{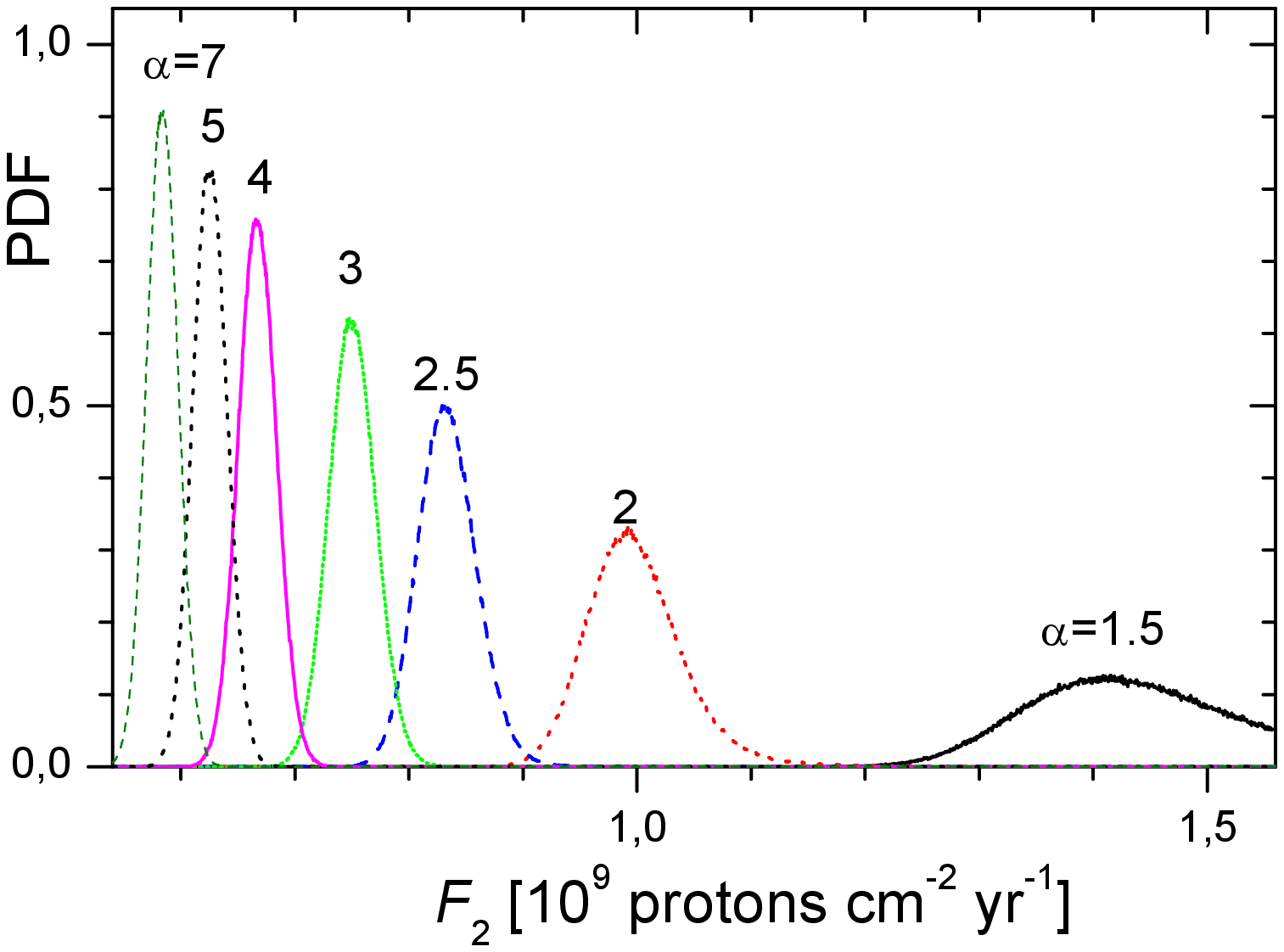}}
\end{center}
\caption{Probability distribution function (PDF) of the values of the average annual fluence ($>30$ MeV) $F_2$ for
 high-fluence events obtained for $10^6$ Monte-Carlo realizations for the $^{14}$C nuclide
  using the power law model (see text).
 Different curves correspond to different values of the power-law spectral index $\alpha$,
 as indicated above each curve.
}
\label{Fig:pdf}       % Give a unique label
\end{figure}

The low-fluence ($F\leq F_0$) events were skipped since they are included into the modern statistics $\langle F_1\rangle$,
 so that $F=0$ for $R<(1-P_0)$.
Then the nuclide's decay with the life-time $\tau$ was applied so that the mean flux is defined as
\begin{equation}
F_2 = {1\over\tau}\int_{t=0}^{12\tau}{F(t)\cdot\exp{(-t/\tau)}\cdot dt},
\label{Eq:f*}
\end{equation}
where the integration is done over 12 life times.
As an example, the distribution of the obtained fluence $F_2$ for a given spectral exponent $\alpha$ (the power-law OPDF)
 is shown in Figure~\ref{Fig:pdf} for $^{14}$C, as calculated from $N=10^6$ simulated series.
One can see that the distributions are nearly Guassian with the mean value corresponding to its mathematical expectation.
% ======

\begin{figure}
\begin{center}
\resizebox{\hsize}{!}{\includegraphics{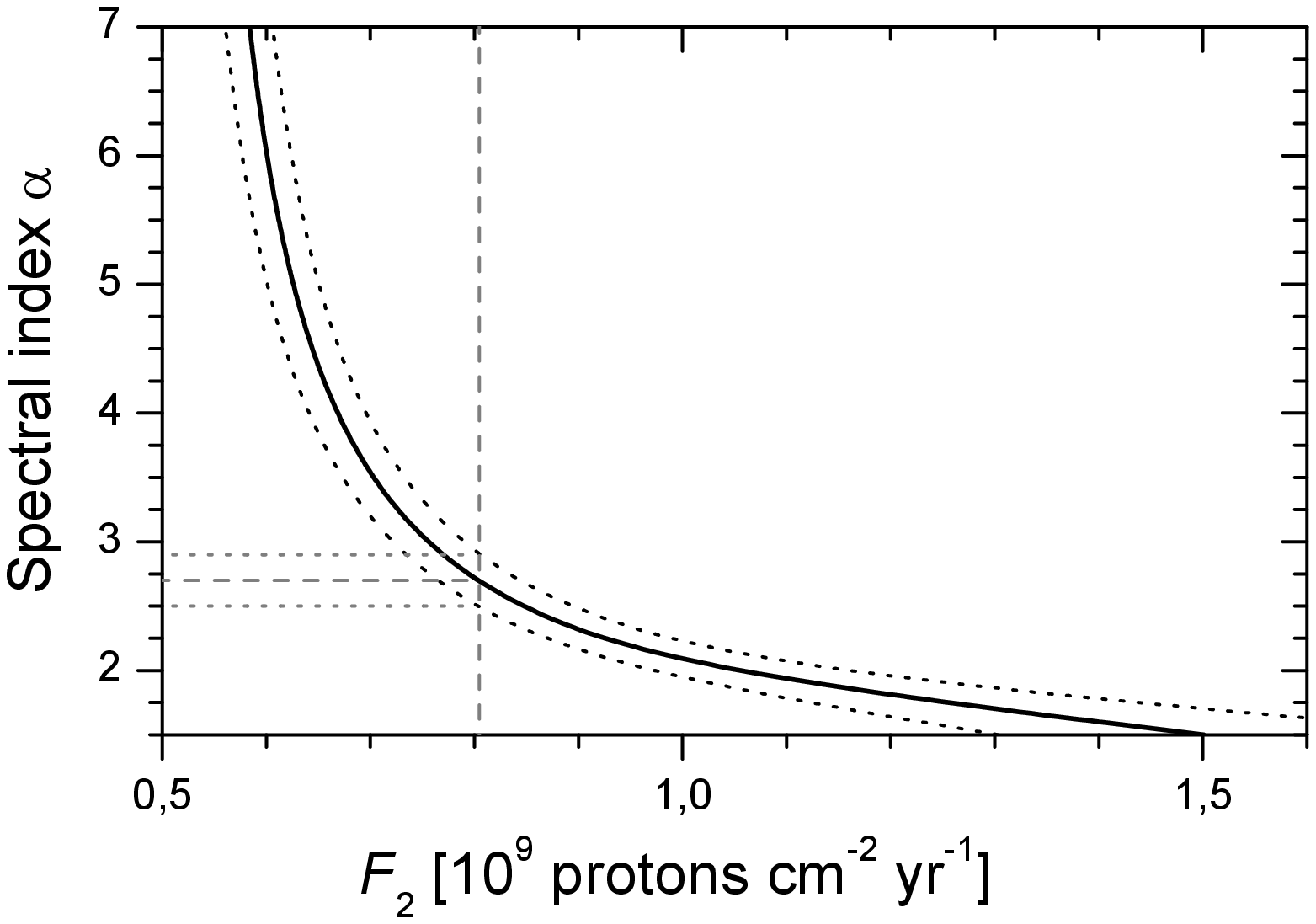}}
\end{center}
\caption{Calibration curve (with its upper and lower 5\% percentiles shown as dotted curves),
 relating the average annual fluence ($>30$ MeV) $\langle F_2 \rangle$ from
 high fluence events to the power-law spectral index $\alpha$ of the high fluence SEP event
 distribution tail for the $^{14}$C nuclide (see text).
The grey dashed lines illustrate how the given $\langle F_2 \rangle$ value is converted into the $\alpha-$values.
}
\label{Fig:calibr}       % Give a unique label
\end{figure}

Now, the spectral exponent $\alpha$ and its uncertainties can be assessed from the measured fluence $F^*$
 for the given nuclide and the above simulations so that
\begin{equation}
F^* - \langle F_1\rangle = \langle F_2(\alpha)\rangle
\label{Eq:eq}
\end{equation}
for the mean value of $\langle F_2 \rangle$ and its upper and lower 5\% percentiles.
The corresponding calibration curve for the power-law OPDF for $^{14}$C is depicted in Figure~\ref{Fig:calibr} to define
 the mean and the upper/lower 5\% percentiles of $\alpha$ from the given value of $\langle F_2 \rangle$.
% ======

As an example, let us consider radiocarbon $^{14}$C measured in lunar rocks.
According to \inlinecite{jull98}, the mean $>30$ MeV fluence reconstructed
 from $^{14}$C lunar record is $F^*=1.33\times 10^9$ protons cm$^{-2}$ yr$^{-1}$.
Considering that $\langle F_1 \rangle=0.52\times 10^9$ protons cm$^{-2}$ yr$^{-1}$, we estimate
 the high-fluence event contribution as $\langle F_2 \rangle=0.81\times 10^9$ protons cm$^{-2}$ yr$^{-1}$.
Using the calibration curve, as illustrated in Figure~\ref{Fig:calibr}, one can
 obtain that the best-fit power-law exponent is $\alpha=2.64\pm0.21$ within the 90\% confidence interval.
This value goes to the sixth column of the first row of Table~\ref{Table:1}.
Similar estimate for the exponential OPDF gives, for $^{14}$C,
 the spectral index $\beta=(0.328\pm0.037)\times 10^{-9}$ cm$^2$ yr.
Similar calculations were done for all other radionuclides for both the OPDF models.

%======================
\begin{table}
\caption{Assessments of the parameters of OPDF from different cosmogenic radionuclide data in lunar rocks.
 Columns correspond to the nuclide, reference to the original data,
 the measured mean annual fluence $F^*$ ($10^9$ protons cm$^{-2}$ yr$^{-1}$), and the corresponding best-fit parameters $\alpha$
 and $\beta$ ($10^{-9}$ cm$^2$ yr) with the 90\% confidence intervals (see text).}
\begin{tabular}{llp{3.2cm}|c|cc}
\hline
\# & Nuclide & Reference & $F^*$ & $\alpha$ & $\beta$\\
\hline
1 & $^{14}$C   & \cite{jull98} & 1.33 & $2.64\pm0.21$ &	$0.328\pm 0.037$\\
2 & $^{41}$Ca  &\cite{fink98}	&	1.77	& $1.67\pm 0.03$ &	$0.134\pm 0.002$ \\
3 & $^{81}$Kr  & \cite{reedy99}	&	1.51	& $2.01\pm 0.02$ &	$0.202\pm 0.003$ \\
4 & $^{36}$Cl  & \cite{nishiizumi09}	&	1.45	& $2.16\pm 0.02$ &	$0.232\pm 0.003$ \\
5 & $^{26}$Al  & \cite{kohl78}	&	0.79	& N/A &	N/A \\
6 & $^{26}$Al  & \cite{grismore01} & 1.74 & $1.69\pm 0.01$ & $0.137\pm 0.001$ \\
7 & $^{10}$Be/$^{26}$Al	& \cite{nishiizumi88}	& 1.10 & $6.93\pm 0.14$ & $1.19\pm 0.03$ \\
8 & $^{10}$Be/$^{26}$Al	& \cite{michel96}	&	0.76	& N/A &	N/A \\
9 & $^{10}$Be/$^{26}$Al	& \cite{fink98}	&	1.01	& N/A &	N/A \\
10& $^{10}$Be/$^{26}$Al	& \cite{nishiizumi09}	&	0.76	& N/A &	N/A \\
11& $^{53}$Mn & \cite{kohl78}	&	0.79	& N/A &	N/A \\
\hline
\end{tabular}
\label{Table:1}
\end{table}

\section{Results and Discussion}

The results of fitting data from different nuclides are shown in Table~\ref{Table:1} for
 the two considered OPDF shapes - power law and exponential ones.
The corresponding distribution are also shown in Figure~\ref{Fig:dist}.

\begin{figure}
\begin{center}
\resizebox{\hsize}{!}{\includegraphics{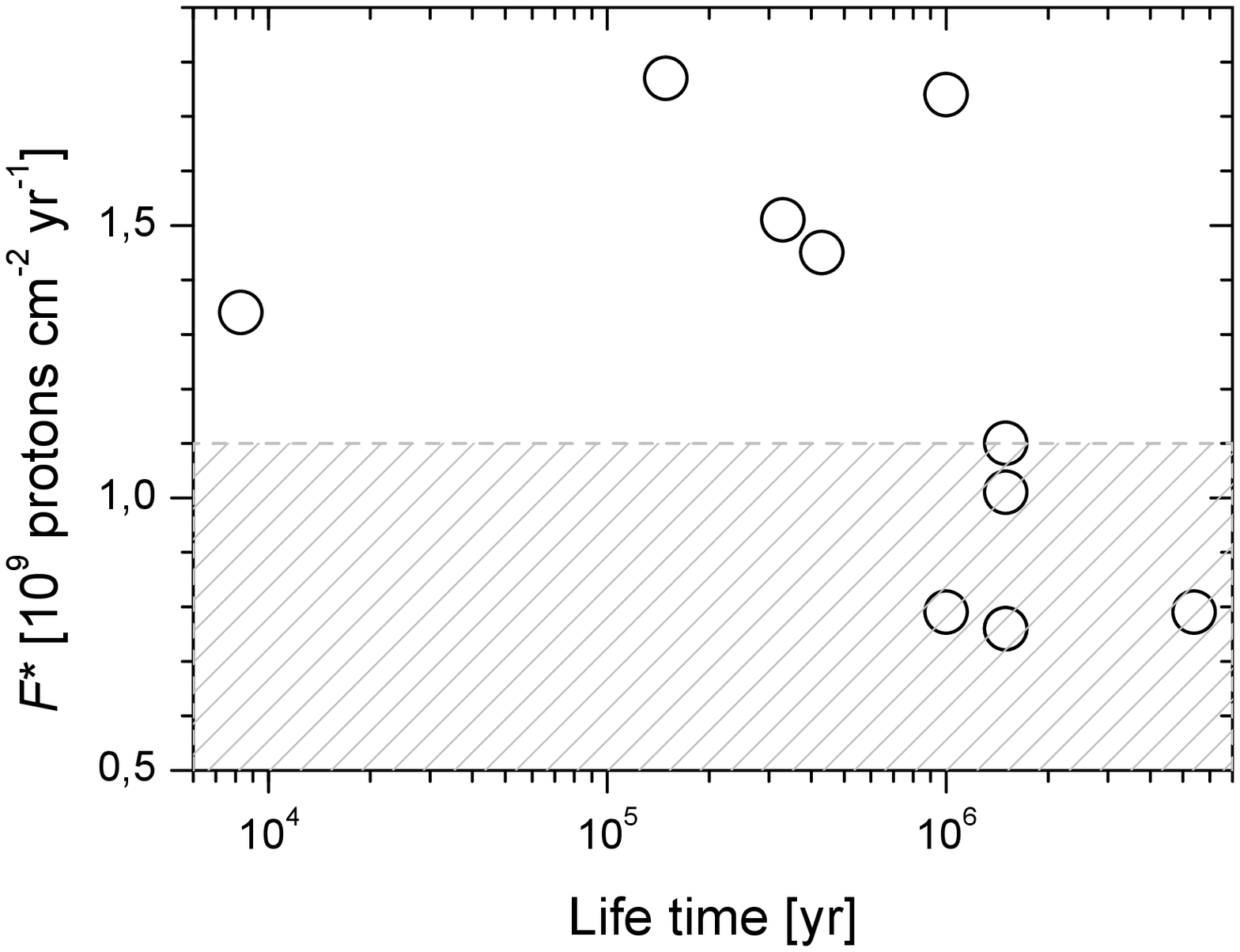}}
\end{center}
\caption{The measured averaged fluences of protons ($>30$ MeV) $F^\ast$ as function of nuclide's life time
 (see Table~1).
 The hatched area corresponds to the averaged fluence $F^\ast$ inconsistent with the measurements of 1955 -- 2007
  and cannot be fit by the model.
}
\label{Fig:life}       % Give a unique label
\end{figure}

All the data from radionuclides with life-time shorter than 0.5 Myr (lines 1--4 in Table~\ref{Table:1})
 yield reasonable results for the OPDF tail.
On the other hand, as we now show the values of $F^\ast < 1.1 \times 10^9$ protons cm$^{-2}$ yr$~{-1}$
 cannot be fitted by either models.
The relation between the estimated fluence and the life time is shown in Figure~\ref{Fig:life}.
Most long-living nuclides, except for the $^{26}$Al-based estimate by \inlinecite{grismore01}, yield
 a low fluence, lower than the recent measurements (shown as the grey dashed line).
In fact, these data cannot be consistent with the present model.
This covers most of the data related to long-living nuclides (lines 5--11 in Table~\ref{Table:1}).

The discrepancy between the results based on short- ($\tau < 0.5$ My) and long- ($\tau =$1--5 Myr)
 living nuclides cannot be ascribed to difference between the measured samples of lunar rocks, as some of the
 data were obtained from the same samples (see Table~3 in \opencite{nishiizumi09}).
Thus, this discrepancy is systematic and can be interpreted in different ways.
One is that the SEP flux was as high as during the modern times for the last 0.5 Myr, but
 was significantly and systematically lower before that.
However, this would imply a dramatic and sharp transition by a factor greater than 2--3
 between the two modes to happen at about a Myr ago, which sounds unrealistic \cite{nishiizumi09}.
Another option is a systematic error in the evaluation of the SEP fluence from lunar samples
 which is accumulated over the time leading to underestimate of the fluence in the far past.
However, studying this kind of uncertainties, {\it e.g.}, correction for erosion or better nuclear cross-sections used in the
 modelling, is beyond the scope of this work.
Accordingly we consider only shorter-living ($\tau < 0.5$ Myr) radionuclides here, stating that OPDF cannot be
 evaluated from long-living nuclide data.

% ======

Next we compare the OPDF obtained from terrestrial cosmogenic radionuclides \cite{usoskin_ApJ_12}
 with those presented here for lunar samples.
The comparison is shown in Figure~\ref{Fig:dist}.
For the power-law OPDF tail (Figure~\ref{Fig.dist}a) one can see that only the $^{14}$C-based lunar
 results are barely consistent with the terrestrial data.
Namely, the data based on lunar $^{14}$C (line 1 in Table~\ref{Table:1}) imply, in the
 framework of the power-law OPDF, that events with the fluence $> 50 \times 10^9$ protons cm$^{-2}$ yr$^{-1}$
  would have occurred, on average, every 5000 yr.
Thus, a few such events would have occurred during the Holocene, each being stronger than the
 greatest observed event of AD775 \cite{miyake12,usoskin_775_13}.
However, so strong events cannot be missed in the terrestrial radionuclide data \cite{usoskin_ApJ_12},
 and the probability that purely randomly no such events occur during the eleven millennia of the
 Holocene is about $0.11$.
On the other hand, a simple $\chi^2$-test suggests that this OPDF tail
 does not fit the terrestrial radionuclide data (open symbols) at the significance level of 0.01.
Thus, the null hypothesis shall be rejected and this OPDF tail cannot be considered as
 consistent with the terrestrial cosmogenic radionuclide data.
We note that the effective time scale covered by $^{14}$C in lunar rocks coincides with
 the Holocene, and thus this data set can be directly compared to terrestrial data.
All other lunar-based radionuclides are in obvious contradiction with the terrestrial data.
Namely, the data based on lunar $^{36}$Cl (line 4 in Table~\ref{Table:1}) imply that events with
 $F^\ast > 50 \times 10^9$ protons cm$^{-2}$ yr$^{-1}$ would have occurred, on average, every 1500 yr.
This leads to the probability that purely randomly no events occur during the
 Holocene, of $\approx 10^{-3}$.
Thus, the power-law OPDF tail is inconsistent with the terrestrial cosmogenic nuclide data at the level of $10^{-3}$.
Other results (lines 3,4 and 5) overestimate the OPDF even greater.
The too steep power-law tail implied by the $^{10}$Be/$^{26}$Al ratio (line 7 in the Table) also is inconsistent with the
 observed data but heavily underestimating the OPDF.
Concluding, the power-law shape of the OPDF tail does not agree with the terrestrial data.

On the other hand, a similar analysis of the exponential shape of the OPDF (Figure~\ref{Fig:dist}b) suggests that the result
 for $^{14}$C (line 1 of Table~\ref{Table:1})
 is well consistent with the terrestrial data.
The null hypothesis that this tail is the same as the measured OPDF for terrestrial cosmogenic nuclide data,
 cannot be rejected (the significance level 0.26).
The other exponential tails, while giving a formally worse fit, still yield a reasonable agreement with the terrestrial data.

\begin{figure}
\begin{center}
\resizebox{\hsize}{!}{\includegraphics{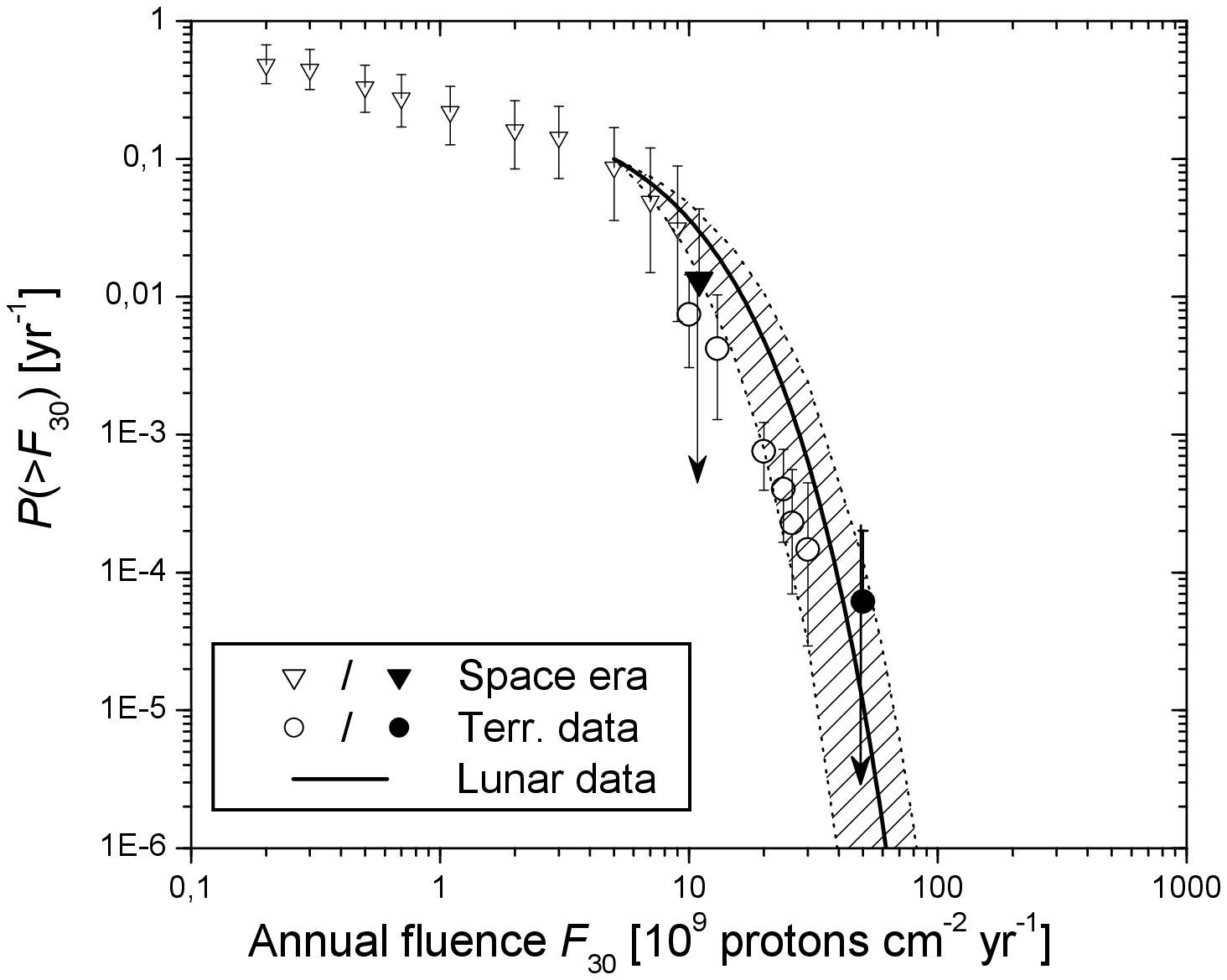}}
\end{center}
\caption{The cumulative occurrence probability distribution of strong SEP events -- a combined plot for time
 scales from yearly to a million of years.
 Triangles represent observations of the space era.
 Circles represent data obtained from terrestrial cosmogenic radionuclide data (Usoskin and Kovaltsov, 2012).
 Filled symbols correspond to the upper limits based on the fact that no events greater than the given fluence value
  have been observed.
 The solid curve with the hatched range (90 \% confidence interval) is the best-fit estimate (this work) based on cosmogenic
  radionuclides with life-time shorter than 0.5 Myr measured on lunar samples.
}
\label{Fig:final}       % Give a unique label
\end{figure}

Since estimates based on individual radionuclides differ from each other quite a bit, and they are somewhat
 uncertain, we also provide a combined estimate of the OPDF for all the nuclides with the life time shorter
 than 0.5 Myr ({\it viz.} lines 1--4 in Table~\ref{Table:1}).
The mean value of $F^\ast$ appears $1.51 \pm 0.18 (\times 10^9$ protons cm$^{-2}$ yr$^{-1}$), for the 90\% confidence interval.
We consider now only the exponential-tail model since the power-law does not fit the terrestrial data as
 described above ({\it cf.} \opencite{nymmik99}).
The corresponding spectral index is found as $\beta=0.202^{+0.122}_{-0.053}$ ($10^{-9}$ cm$^2$ yr).
This best-fit OPDF is shown in Figure~\ref{Fig:final} as the solid curve with the hatched range.
One can see that all the three time scales considered, {\it viz.} years-decades measured during the space era,
 centennia-millennia from terrestrial radionuclides, and the scale of up to a million years,
 are consistent in the OPDF.
They indicate a strong exponential rollover for strong SEP events that is theoretically expected because of the
 effects of the ion-wave interactions leading to the streaming limit of fluxes observed by space-borne
 instruments during large SEP events \cite{reames04}.

% ======

\section{Conclusions}

We have assessed the occurrence probability distribution function for strong SEP events with the
 fluence of $>30$ MeV protons exceeding $5\times 10^9$ (protons cm$^{-2}$ yr$^{-1}$),
 from data of different cosmogenic radionuclides measured in lunar samples.
We present, in Figure~\ref{Fig:final}, a combined cumulative occurrence probability distribution
 of SEP events based on three time scales: directly measured SEP fluences for the last 60 years;
 estimates based on terrestrial cosmogenic radionuclides $^{10}$Be and $^{14}$C for the multi-millennial (Holocene) time scale;
 and cosmogenic radionuclides measured in lunar rocks on the time scale of up to 1 Myr.
All the three time scales yield a consistent distribution.

We conclude that:
\begin{itemize}
\item
All SEP fluences estimated for long-living isotopes with the life time greater than 1 Myr, in lunar rocks, are inconsistent
 with the terrestrial data on decadal to multimillennial time scale.
Accordingly, the average SEP fluxes cannot be reliably assessed on the time scale of longer than 1 Myr.
\item
The data suggests a strong rollover of the occurrence probability so that the SEP events with $F_{30}$ fluence greater than $10^{11}$
 protons cm$^{-2}$ yr$^{-1}$
 are not expected at the Myr time scale.
\item
The best-fit result for the exponential tail of the occurrence probability distribution function [Equation~(\ref{Eq:EXP})]
 yields the value of $\beta$ in the range of 0.15--0.32 ($\times 10^9$ cm$^2$ yr).
\end{itemize}
%

%=======================================
\acknowledgements{
We are grateful to Peggy Shea and Don Smart for data on SEP fluxes.
GA acknowledges partial support from the Program No. 22 of Presidium RAS and from the Academy of Finland.
We thank the International Space Studies Institute in Bern, Switzerland, for
support of the team "Extreme Solar Flares as Drivers of Space Weather".
}

%%% %%%%%%%%%%%%%%%%%%%%%%%%%%%%%%%%%%%%%%%%%%%%%%%%%%%%%%%%%%%
%% Bibliography
%
% Using BibTeX
%
%\bibliographystyle{spr-mp-sola}
%\bibliographystyle{spr-mp-sola-cnd} %% Alternative style: no title, no concluding page
%\bibliography{C:/DATA__/USOSKIN/papers/usoskin_all}
%\bibliography{J:/USOSKIN/papers/usoskin_all}

\begin{thebibliography}{26}
% BibTex style file: spr-mp-sola.bst (nameyear), 2011-09-16
\ifx \bisbn   \undefined \def \bisbn  #1{ISBN #1}\fi
\ifx \binits  \undefined \def \binits#1{#1}\fi
\ifx \bauthor  \undefined \def \bauthor#1{#1}\fi
\ifx \batitle  \undefined \def \batitle#1{#1}\fi
\ifx \bjtitle  \undefined \def \bjtitle#1{\textit{#1}}\fi
\ifx \bvolume  \undefined \def \bvolume#1{\textbf{#1}}\fi
\ifx \byear  \undefined \def \byear#1{#1}\fi
\ifx \bissue  \undefined \def \bissue#1{#1}\fi
\ifx \bfpage  \undefined \def \bfpage#1{#1}\fi
\ifx \blpage  \undefined \def \blpage #1{#1}\fi
\ifx \burl  \undefined \def \burl#1{\textsf{#1}}\fi
\ifx \href  \undefined \def \href#1#2{\textsf{#2}}\fi
\ifx \doiurl  \undefined \def
  \doiurl#1{\href{http://dx.doi.org/#1}{\textsf{#1}}}\fi
\ifx \betal  \undefined \def \betal{\textit{et al.}}\fi
\ifx \binstitute  \undefined \def \binstitute#1{#1}\fi
\ifx \bctitle  \undefined \def \bctitle#1{#1}\fi
\ifx \beditor  \undefined \def \beditor#1{#1}\fi
\ifx \bpublisher  \undefined \def \bpublisher#1{#1}\fi
\ifx \bbtitle  \undefined \def \bbtitle#1{\textit{#1}}\fi
\ifx \bedition  \undefined \def \bedition#1{#1}\fi
\ifx \bseriesno  \undefined \def \bseriesno#1{\textbf{#1}}\fi
\ifx \blocation  \undefined \def \blocation#1{#1}\fi
\ifx \bsertitle  \undefined \def \bsertitle#1{\textit{#1}}\fi
\ifx \bsnm \undefined \def \bsnm#1{#1}\fi
\ifx \bsuffix \undefined \def \bsuffix#1{#1}\fi
\ifx \bparticle \undefined \def \bparticle#1{#1}\fi
\ifx \barticle \undefined \def \barticle#1{}\fi
\ifx \botherref \undefined \def \botherref#1{}\fi
\ifx \url \undefined \def \url#1{\textsf{#1}}\fi
\ifx \bchapter \undefined \def \bchapter#1{}\fi
\ifx \bbook \undefined \def \bbook#1{}\fi
\ifx \bcomment \undefined \def \bcomment#1{#1}\fi
\ifx \oauthor \undefined \def \oauthor#1{#1}\fi
\ifx \citeauthoryear \undefined \def \citeauthoryear#1{#1}\fi
\def \endbibitem {}
\ifx \bconflocation  \undefined \def \bconflocation#1{#1} \fi

\bibitem[\protect\citeauthoryear{Fink \textit{et~al.}}{1998}]{fink98}
\begin{barticle}
\bauthor{\bsnm{Fink}, \binits{D.}},
\bauthor{\bsnm{Klein}, \binits{J.}},
\bauthor{\bsnm{Middleton}, \binits{R.}},
\bauthor{\bsnm{Vogt}, \binits{S.}},
\bauthor{\bsnm{Herzog}, \binits{G.F.}},
\bauthor{\bsnm{Reedy}, \binits{R.C.}}:
\byear{1998},
\batitle{$^{41}$Ca, $^{26}$Al, and $^{10}$Be in lunar basalt 74275 and $^{10}$Be in the double
  drive tube 74002/74001}.
\bjtitle{Geochim. Cosmochim. Acta}
\bvolume{62},
\bfpage{2389}\,--\,\blpage{2402}.
doi:\doiurl{10.1016/S0016-7037(98)00134-3}.
\end{barticle}
\endbibitem

\bibitem[\protect\citeauthoryear{{Grismore} \textit{et~al.}}{2001}]{grismore01}
\begin{barticle}
\bauthor{\bsnm{{Grismore}}, \binits{R.}},
\bauthor{\bsnm{{Llewellyn}}, \binits{R.A.}},
\bauthor{\bsnm{{Brown}}, \binits{M.D.}},
\bauthor{\bsnm{{Dowson}}, \binits{S.T.}},
\bauthor{\bsnm{{Cumblidge}}, \binits{K.}}:
\byear{2001},
\batitle{{Measurements of the concentrations of $^{26}$Al in lunar rocks 15555
  and 60025}}.
\bjtitle{Earth Planet. Sci. Lett.}
\bvolume{187},
\bfpage{163}\,--\,\blpage{171}.
doi:\doiurl{10.1016/S0012-821X(01)00271-0}.
\end{barticle}
\endbibitem

\bibitem[\protect\citeauthoryear{Hudson}{{2010}}]{hudson10}
\begin{barticle}
\bauthor{\bsnm{Hudson}, \binits{H.S.}}:
\byear{{2010}},
\batitle{{Solar flares add up}}.
\bjtitle{{Nature Phys.}}
\bvolume{{6}},
\bfpage{637}\,--\,\blpage{638}.
doi:\doiurl{{10.1038/nphys1764}}.
\end{barticle}
\endbibitem

\bibitem[\protect\citeauthoryear{Jull \textit{et~al.}}{1998}]{jull98}
\begin{barticle}
\bauthor{\bsnm{Jull}, \binits{A.J.T.}},
\bauthor{\bsnm{Cloudt}, \binits{S.}},
\bauthor{\bsnm{Donahue}, \binits{D.J.}},
\bauthor{\bsnm{Sisterson}, \binits{J.M.}},
\bauthor{\bsnm{Reedy}, \binits{R.C.}},
\bauthor{\bsnm{Masarik}, \binits{J.}}:
\byear{1998},
\batitle{$^{14}$C depth profiles in Apollo 15 and 17 cores and lunar rock 68815}.
\bjtitle{Geochim. Cosmochim. Acta}
\bvolume{62},
\bfpage{3025}\,--\,\blpage{3036}.
doi:\doiurl{10.1016/S0016-7037(98)00193-8}.
\end{barticle}
\endbibitem

\bibitem[\protect\citeauthoryear{{Jun} \textit{et~al.}}{2007}]{jun07}
\begin{barticle}
\bauthor{\bsnm{{Jun}}, \binits{I.}},
\bauthor{\bsnm{{Swimm}}, \binits{R.T.}},
\bauthor{\bsnm{{Ruzmaikin}}, \binits{A.}},
\bauthor{\bsnm{{Feynman}}, \binits{J.}},
\bauthor{\bsnm{{Tylka}}, \binits{A.J.}},
\bauthor{\bsnm{{Dietrich}}, \binits{W.F.}}:
\byear{2007},
\batitle{{Statistics of solar energetic particle events: Fluences, durations,
  and time intervals}}.
\bjtitle{Adv. Space Res.}
\bvolume{40},
\bfpage{304}\,--\,\blpage{312}.
doi:\doiurl{10.1016/j.asr.2006.12.019}.
\end{barticle}
\endbibitem

\bibitem[\protect\citeauthoryear{Kohl \textit{et~al.}}{1978}]{kohl78}
\begin{bchapter}
\bauthor{\bsnm{Kohl}, \binits{C.P.}},
\bauthor{\bsnm{Murrell}, \binits{M.T.}},
\bauthor{\bsnm{Russ~III}, \binits{G.P.}},
\bauthor{\bsnm{Arnold}, \binits{J.R.}}:
\byear{1978},
\bctitle{Evidence for the constancy of the solar cosmic ray flux over the past
  ten million years: $^{53}$Mn and $^{26}$Al measurements}.
In: \bbtitle{Proceeding of the Ninth Lunar and Planetary Science Conference},
\bsertitle{Geochim. Cosmochim. Acta Suppl.}
\bseriesno{10},
\bpublisher{Pergamon Press},
\blocation{New York},
\bfpage{2299}\,--\,\blpage{2310}.
\end{bchapter}
\endbibitem

\bibitem[\protect\citeauthoryear{McCracken \textit{et~al.}}{2001}]{mccracken01}
\begin{barticle}
\bauthor{\bsnm{McCracken}, \binits{K.G.}},
\bauthor{\bsnm{Dreschhoff}, \binits{G.A.M.}},
\bauthor{\bsnm{Zeller}, \binits{E.J.}},
\bauthor{\bsnm{Smart}, \binits{D.F.}},
\bauthor{\bsnm{Shea}, \binits{M.A.}}:
\byear{2001},
\batitle{Solar cosmic ray events for the period 1561-1994: 1. Identification in
  polar ice, 1561-1950}.
\bjtitle{J. Geophys. Res.}
\bvolume{106},
\bfpage{21585}\,--\,\blpage{21598}.
doi:\doiurl{10.1029/2000JA000237}.
\end{barticle}
\endbibitem

\bibitem[\protect\citeauthoryear{Michel, Leya, and Borges}{1996}]{michel96}
\begin{barticle}
\bauthor{\bsnm{Michel}, \binits{R.}},
\bauthor{\bsnm{Leya}, \binits{I.}},
\bauthor{\bsnm{Borges}, \binits{L.}}:
\byear{1996},
\batitle{Production of cosmogenic nuclides in meteoroids: accelerator
  experiments and model calculations to decipher the cosmic ray record in
  extraterrestrial matter.}
\bjtitle{Nucl. Inst. Meth. Phys. Res. B}
\bvolume{113},
\bfpage{434}\,--\,\blpage{444}.
doi:\doiurl{10.1016/0168-583X(95)01345-8}.
\end{barticle}
\endbibitem

\bibitem[\protect\citeauthoryear{Miyake \textit{et~al.}}{2012}]{miyake12}
\begin{barticle}
\bauthor{\bsnm{Miyake}, \binits{F.}},
\bauthor{\bsnm{Nagaya}, \binits{K.}},
\bauthor{\bsnm{Masuda}, \binits{K.}},
\bauthor{\bsnm{Nakamura}, \binits{T.}}:
\byear{2012},
\batitle{{A signature of cosmic-ray increase in AD 774–775 from tree rings in Japan}}.
\bjtitle{Nature}
\bvolume{486},
\bfpage{240}\,--\,\blpage{242}.
doi:\doiurl{10.1038/nature11123}.
\end{barticle}
\endbibitem

\bibitem[\protect\citeauthoryear{{Nishiizumi}
  \textit{et~al.}}{1988}]{nishiizumi88}
\begin{bchapter}
\bauthor{\bsnm{{Nishiizumi}}, \binits{K.}},
\bauthor{\bsnm{{Imamura}}, \binits{M.}},
\bauthor{\bsnm{{Kohl}}, \binits{C.P.}},
\bauthor{\bsnm{{Nagai}}, \binits{H.}},
\bauthor{\bsnm{{Kobayashi}}, \binits{K.}},
\bauthor{\bsnm{{Yoshida}}, \binits{K.}},
\bauthor{\bsnm{{Yamashita}}, \binits{H.}},
\bauthor{\bsnm{{Reedy}}, \binits{R.C.}},
\bauthor{\bsnm{{Honda}}, \binits{M.}},
\bauthor{\bsnm{{Arnold}}, \binits{J.R.}}:
\byear{1988},
\bctitle{{$^{10}$Be profiles in lunar surface rock 68815}}.
In: \beditor{\bsnm{{Ryder}}, \binits{G.}} (ed.)
\bsertitle{Proceeding of the Eighteenth Lunar and Planetary Science Conference},
\bpublisher{Cambridge University Press},
\blocation{Cambridge},
\bfpage{79}\,--\,\blpage{85}.
\end{bchapter}
\endbibitem

\bibitem[\protect\citeauthoryear{{Nishiizumi}
  \textit{et~al.}}{2009}]{nishiizumi09}
\begin{barticle}
\bauthor{\bsnm{{Nishiizumi}}, \binits{K.}},
\bauthor{\bsnm{{Arnold}}, \binits{J.R.}},
\bauthor{\bsnm{{Kohl}}, \binits{C.P.}},
\bauthor{\bsnm{{Caffee}}, \binits{M.W.}},
\bauthor{\bsnm{{Masarik}}, \binits{J.}},
\bauthor{\bsnm{{Reedy}}, \binits{R.C.}}:
\byear{2009},
\batitle{{Solar cosmic ray records in lunar rock 64455}}.
\bjtitle{Geochim. Cosmochim. Acta}
\bvolume{73},
\bfpage{2163}\,--\,\blpage{2176}.
doi:\doiurl{10.1016/j.gca.2008.12.021}.
\end{barticle}
\endbibitem

\bibitem[\protect\citeauthoryear{{Nymmik}}{1999}]{nymmik99}
\begin{bchapter}
\bauthor{\bsnm{{Nymmik}}, \binits{R.}}:
\byear{1999},
\bctitle{{SEP event distribution function as inferred from spaceborne measurements
and lunar rock isotopic data}}.
In: Kieda, D., Salamon, M., Dingus, B. (eds.),
\bbtitle{26th Proc. 26th Int. Cosmic Ray Conf.}
\bseriesno{6},
\bfpage{268}\,--\,\blpage{271}.
\end{bchapter}
\endbibitem

\bibitem[\protect\citeauthoryear{{Reames}}{2004}]{reames04}
\begin{barticle}
\bauthor{\bsnm{{Reames}}, \binits{D.V.}}:
\byear{2004},
\batitle{{Solar energetic particle variations}}.
\bjtitle{Adv. Space Res.}
\bvolume{34},
\bfpage{381}\,--\,\blpage{390}.
doi:\doiurl{10.1016/j.asr.2003.02.046}.
\end{barticle}
\endbibitem

\bibitem[\protect\citeauthoryear{{Reedy}}{1999}]{reedy99}
\begin{bchapter}
\bauthor{\bsnm{{Reedy}}, \binits{R.C.}}:
\byear{1999},
\bctitle{{Variations in solar-proton fluxes over the last million years}}.
In: \bbtitle{Lunar and Planetary Institute Science Conference Abstracts}
\bseriesno{30},
\bfpage{1643}.
\end{bchapter}
\endbibitem

\bibitem[\protect\citeauthoryear{Reedy}{1996}]{reedy96}
\begin{bchapter}
\bauthor{\bsnm{Reedy}, \binits{R.C.}}:
\byear{1996},
\bctitle{Constraints on solar particle events from comparisons of recent events
  and million-year averages}.
In: \beditor{\bsnm{Balasubramaniam}, \binits{K.S.}},
\beditor{\bsnm{Keil}, \binits{S.L.}},
\beditor{\bsnm{Smartt}, \binits{R.N.}} (eds.)
\bbtitle{Solar Drivers of the Interplanetary and Terrestrial Disturbances},
\bsertitle{ASP Conf. Ser.}
\bseriesno{95},
\bconflocation{San Francisco, U.S.A.},
\bfpage{429}\,--\,\blpage{436}.
\end{bchapter}
\endbibitem

\bibitem[\protect\citeauthoryear{{Schrijver}
  \textit{et~al.}}{2012}]{schrijver12}
\begin{barticle}
\bauthor{\bsnm{{Schrijver}}, \binits{C.J.}},
\bauthor{\bsnm{Beer}, \binits{J.}},
\bauthor{\bsnm{Baltensperger}, \binits{U.}},
\bauthor{\bsnm{Cliver}, \binits{E.W.}},
\bauthor{\bsnm{G\"udel}, \binits{M.}},
\bauthor{\bsnm{Hudson}, \binits{H.S.}},
\bauthor{\bsnm{McCracken}, \binits{K.G.}},
\bauthor{\bsnm{Osten}, \binits{R.A.}},
\bauthor{\bsnm{Peter}, \binits{T.}},
\bauthor{\bsnm{Soderblom}, \binits{D.R.}},
\bauthor{\bsnm{Usoskin}, \binits{I.G.}},
\bauthor{\bsnm{Wolff}, \binits{E.W.}}:
\byear{2012},
\batitle{{Estimating the frequency of extremely energetic solar events, based
  on solar, stellar, lunar, and terrestrial records}}.
\bjtitle{J. Geophys. Res.}
\bvolume{117},
\bfpage{{A08103}}.
doi:\doiurl{10.1029/2012JA017706}.
\end{barticle}
\endbibitem

\bibitem[\protect\citeauthoryear{{Shea} and {Smart}}{2012}]{shea12}
\begin{barticle}
\bauthor{\bsnm{{Shea}}, \binits{M.A.}},
\bauthor{\bsnm{{Smart}}, \binits{D.F.}}:
\byear{2012},
\batitle{{Space weather and the ground-level solar proton events of the 23rd solar cycle}}.
\bjtitle{Space Sci. Rev.}
\bvolume{171},
\bfpage{161}\,--\,\blpage{188}.
doi:\doiurl{10.1007/s11214-012-9923-z}.
\end{barticle}
\endbibitem

\bibitem[\protect\citeauthoryear{Shea and Smart}{1990}]{shea90}
\begin{barticle}
\bauthor{\bsnm{Shea}, \binits{M.A.}},
\bauthor{\bsnm{Smart}, \binits{D.F.}}:
\byear{1990},
\batitle{A summary of major solar proton events}.
\bjtitle{Solar Phys.}
\bvolume{127},
\bfpage{297}\,--\,\blpage{320}.
doi:\doiurl{10.1007/BF00152170}.
\end{barticle}
\endbibitem

\bibitem[\protect\citeauthoryear{Shea \textit{et~al.}}{2006}]{shea06}
\begin{barticle}
\bauthor{\bsnm{Shea}, \binits{M.A.}},
\bauthor{\bsnm{Smart}, \binits{D.F.}},
\bauthor{\bsnm{McCracken}, \binits{K.G.}},
\bauthor{\bsnm{Dreschhoff}, \binits{G.A.M.}},
\bauthor{\bsnm{Spence}, \binits{H.E.}}:
\byear{2006},
\batitle{Solar proton events for 450 years: The carrington event in
  perspective}.
\bjtitle{Adv. Space Res.}
\bvolume{38},
\bfpage{232}\,--\,\blpage{238}.
doi:\doiurl{10.1016/j.asr.2005.02.100}.
\end{barticle}
\endbibitem

\bibitem[\protect\citeauthoryear{{Smart} \textit{et~al.}}{2006}]{smart06}
\begin{barticle}
\bauthor{\bsnm{{Smart}}, \binits{D.F.}},
\bauthor{\bsnm{{Shea}}, \binits{M.A.}},
\bauthor{\bsnm{{Spence}}, \binits{H.E.}},
\bauthor{\bsnm{{Kepko}}, \binits{L.}}:
\byear{2006},
\batitle{Two groups of extremely large $>$30 MeV solar proton fluence events}.
\bjtitle{Adv. Space Res.}
\bvolume{37},
\bfpage{1734}\,--\,\blpage{1740}.
doi:\doiurl{10.1016/j.asr.2005.09.008}.
\end{barticle}
\endbibitem

\bibitem[\protect\citeauthoryear{{Thomas} \textit{et~al.}}{2013}]{thomas13}
\begin{botherref}
\oauthor{\bsnm{{Thomas}}, \binits{B.C.}},
\oauthor{\bsnm{{Melott}}, \binits{A.L.}},
\oauthor{\bsnm{{Arkenberg}}, \binits{K.R.}},
\oauthor{\bsnm{{Snyder}}, \binits{B.R.} \bsuffix{II}}:
\byear{2013},
\batitle{Terrestrial effects of possible astrophysical sources of an AD 774-775
  increase in 14C production}.
\bjtitle{Geophys. Res. Lett.}
\bvolume{40},
\bfpage{1237}\,--\,\blpage{1240}.
doi:\doiurl{10.1002/grl.50222}.
\end{botherref}
\endbibitem

\bibitem[\protect\citeauthoryear{{Usoskin} and
  {Kovaltsov}}{2012}]{usoskin_ApJ_12}
\begin{barticle}
\bauthor{\bsnm{{Usoskin}}, \binits{I.G.}},
\bauthor{\bsnm{{Kovaltsov}}, \binits{G.A.}}:
\byear{2012},
\batitle{{Occurrence of extreme solar particle events: Assessment from historical proxy data}}.
\bjtitle{Astrophys. J.}
\bvolume{757},
\bfpage{92}.
doi:\doiurl{10.1088/0004-637X/757/1/92}.
\end{barticle}
\endbibitem

\bibitem[\protect\citeauthoryear{{Usoskin}
  \textit{et~al.}}{2013}]{usoskin_775_13}
\begin{barticle}
\bauthor{\bsnm{{Usoskin}}, \binits{I.G.}},
\bauthor{\bsnm{{Kromer}}, \binits{B.}},
\bauthor{\bsnm{{Ludlow}}, \binits{F.}},
\bauthor{\bsnm{{Beer}}, \binits{J.}},
\bauthor{\bsnm{{Friedrich}}, \binits{M.}},
\bauthor{\bsnm{{Kovaltsov}}, \binits{G.A.}},
\bauthor{\bsnm{{Solanki}}, \binits{S.K.}},
\bauthor{\bsnm{{Wacker}}, \binits{L.}}:
\byear{2013},
\batitle{{The AD775 cosmic event revisited: the Sun is to blame}}.
\bjtitle{Astron. Astrophys.}
\bvolume{552},
\bfpage{{L3}}.
doi:\doiurl{10.1051/0004-6361/201321080}.
\end{barticle}
\endbibitem

\bibitem[\protect\citeauthoryear{Usoskin
  \textit{et~al.}}{2006}]{usoskin_GRL_SCR06}
\begin{barticle}
\bauthor{\bsnm{Usoskin}, \binits{I.G.}},
\bauthor{\bsnm{Solanki}, \binits{S.K.}},
\bauthor{\bsnm{Kovaltsov}, \binits{G.A.}},
\bauthor{\bsnm{Beer}, \binits{J.}},
\bauthor{\bsnm{Kromer}, \binits{B.}}:
\byear{2006},
\batitle{Solar proton events in cosmogenic isotope data}.
\bjtitle{Geophys. Res. Lett.}
\bvolume{33},
\bfpage{L08107}.
doi:\doiurl{10.1029/2006GL026059}.
\end{barticle}
\endbibitem

\bibitem[\protect\citeauthoryear{Vogt, Herzog, and Reedy}{1990}]{vogt90}
\begin{barticle}
\bauthor{\bsnm{Vogt}, \binits{S.}},
\bauthor{\bsnm{Herzog}, \binits{G.F.}},
\bauthor{\bsnm{Reedy}, \binits{R.C.}}:
\byear{1990},
\batitle{Cosmogenic nuclides in extraterrestrial materials}.
\bjtitle{Rev. Geophys.}
\bvolume{28},
\bfpage{253}\,--\,\blpage{275}.
doi:\doiurl{10.1029/RG028i003p00253}.
\end{barticle}
\endbibitem

\bibitem[\protect\citeauthoryear{{Wolff} \textit{et~al.}}{2012}]{wolff12}
\begin{barticle}
\bauthor{\bsnm{{Wolff}}, \binits{E.W.}},
\bauthor{\bsnm{{Bigler}}, \binits{M.}},
\bauthor{\bsnm{{Curran}}, \binits{M.A.J.}},
\bauthor{\bsnm{{Dibb}}, \binits{J.E.}},
\bauthor{\bsnm{{Frey}}, \binits{M.M.}},
\bauthor{\bsnm{{Legrand}}, \binits{M.}},
\bauthor{\bsnm{{McConnell}}, \binits{J.R.}}:
\byear{2012},
\batitle{{The Carrington event not observed in most ice core nitrate records}}.
\bjtitle{Geophys. Res. Lett.}
\bvolume{39},
\bfpage{{L08503}}.
doi:\doiurl{10.1029/2012GL051603}.
\end{barticle}
\endbibitem

\end{thebibliography}
%
% Without BibTeX

\end{article}
\end{document}